# Emergence of Exceptional Points in Periodic Metastructures with Hidden PT-symmetric Defects

**Yanghao Fang**
Department of Materials Science and Engineering,
University of Wisconsin-Madison, Madison, WI 53706
e-mail: fang62@wisc.edu

**Tsampikos Kottos**
Wave Transport in Complex Systems Lab, Department of Physics,
Wesleyan University, Middletown, CT 06459
e-mail: tkottos@wesleyan.edu

**Ramathasan Thevamaran[1]**
Department of Engineering Physics,
University of Wisconsin-Madison, Madison, WI 53706
e-mail: thevamaran@wisc.edu

## ABSTRACT

*We study the elastodynamics of a periodic metastructure incorporating a defect pair that enforces a parity-time (PT) symmetry due to a judiciously engineered imaginary impedance elements— one having energy amplification (gain) and the other having an equivalent attenuation (loss) mechanism. We show that their presence affects the initial band structure of the periodic Hermitian metastructure and leads to the formation of numerous exceptional points (EPs) which are mainly located at the band edges where the local density of modes is higher. The spatial location of the PT-symmetric defect serves as an additional control over the number of emerging EPs in the corresponding spectra as well as the critical non-Hermitian (gain/loss) strength required to create the first EP—a specific defect location minimizes the critical non-Hermitian strength. We use both finite element and coupled-mode-theory-based models to investigate these metastructures, and use a time-independent second-order perturbation theory to further*

*demonstrate the influence of the size of the metastructure and the PT-symmetric defect location on the*

*minimum non-Hermitian strength required to create the first EP in a band. Our findings motivate feasible*

*designs for the experimental realization of EPs in elastodynamic metastructures.*

*Keywords: PT-symmetry, exceptional point, elastodynamics, defect, metastructure*

**INTRODUCTION**

The parity-time (PT) symmetry, unlike the apparent geometric symmetries such as the translational and rotational symmetries found in common engineering designs, is hidden in the equations of motion that describe the dynamical system. Its realization requires judiciously designed balanced energy amplification (gain) and attenuation (loss) mechanisms and impedance profile of the metastructure. A PT-symmetric system is described by a non-Hermitian Hamiltonian which remains invariant under a combined PT operation, where P is the parity operator (defined by momentum operator $\hat{p} \rightarrow -\hat{p}$ and position operator $\hat{x} \rightarrow -\hat{x}$) and T is the time operator ($\hat{p} \rightarrow -\hat{p}, \hat{x} \rightarrow \hat{x}, i \rightarrow -i$). Such class of Hamiltonians can possess entirely real spectra which were demonstrated first in theoretical quantum and mathematical physics [1–4] and later explored in other areas including solid states [5], optics and photonics [6–12], microwave [13,14], electronics [15,16], acoustics [17–23], and elastodynamics [24,25]. When a parameter of the PT-symmetric system, e.g., the non-Hermitian strength $\gamma$, reaches a threshold value, a PT-phase-transition point will occur where two or more eigenvalues (see Fig. 1) and the corresponding eigenvectors become degenerate simultaneously. This transition point is a branch-point singularity named exceptional point (EP) which can lead to intriguing wave phenomena such as unidirectional invisibility [12,17,19], shadow-free sensing [18], and wave switching [14,23].

The recent implementation of PT-concepts in elastodynamics has opened up new fascinating opportunities. For example, we have experimentally and theoretically shown the EP-based hypersensitive sensing in elastodynamics [24] where a pair of non-Hermitian pillars that resonate in torsional mode is embedded in an elastic substrate that modulates the coupling between them through its bending modes and provides the ability to detect a growing defect in the substrate. In an electromechanical accelerometer, we have demonstrated that the EP-based sensing scheme can lead to three-fold signal-to-noise enhancement compared to the system operating away from the EP, enabling hypersensitive sensing [26]. Most recently, we have experimentally demonstrated a novel utility of the eigenvector coalescence associated with the EP, which enables enhancing emissivity by a non-Hermitian metamaterial—i.e., the actuation force from an actuator can be enhanced beyond the standard Purcell enhancement by appropriately coupling it to a non-Hermitian metamaterial operating at the proximity of the EP [27]. The potential for the formation of EPs in various Hermitian and non-Hermitian continuum elastodynamic waveguides and layered media have also been proposed theoretically where the non-Hermiticity arises as a function of hybridization of different elastic wave modes [28–30]. By considering quasiperiodic, geometric fractal, and aperiodic elastodynamic systems with PT-symmetry, we have also shown that the emergence of the EPs follows a universal route [25] where (i) the critical non-Hermitian strength $\gamma_{EP}$ (the precise intensity of balanced gain/loss introduced into the system leading to the EP formation) is linearly proportional to the initial split $\Delta_0$ between the coalescing modes (the frequency difference between two modes in the initial Hermitian system which will degenerate and form EP upon introducing sufficient balanced gain/loss), i.e., $\gamma_{EP} \sim \Delta_0$, and (ii) the scale-free distribution of the EPs is directly predicted by the fractal dimension $D$ of the spectra, i.e., the probability density function $\mathcal{P}(\gamma_{EP}) \sim \gamma_{EP}^{-(1+D)}$.

Realizing such elastodynamic systems that support multiple EPs with potential multi-scale sensitivity, however, requires judicious spatio-temporal control of multiple balanced gain and loss mechanisms. This makes their realization challenging because of the external circuitry that controls multiple of those elements individually through piezoelectric coupling. Here, we show that appropriately positioning a pair of defects — with no apparent defects in geometry or periodicity—into Hermitian periodic elastodynamic metastructures can enforce a PT-symmetry that results in numerous EPs that can be exploited for engineering applications with the simplicity of gain/loss control circuitry applied only to those defects. Defects are known to affect the physical properties of common materials. When they embody PT-symmetric characteristics, they offer new possibilities in solitons [31–34], defect states [35], bound states [36], robust localized modes [37], scattering of linear and nonlinear waves in waveguide arrays [38]. We also show that the non-Hermitian strength $\gamma_{EP}$ required for EP formation and the density of EPs that emerge can additionally be tailored as a function of the location and non-Hermitian characteristics of the defect pair. We use finite element (FE) analysis and coupled-mode theory (CMT)-based model combined with time-independent second-order perturbation (TISOP) theory to study the relationship between the location and characteristics of the defect pair and the nature of the first emerging EP in the spectrum. We also compare this metastructure's response to its entire PT-symmetric counterpart, i.e., a PT-symmetric periodic metastructure with half of it having gain and the other half having loss.

**RESULTS AND DISCUSSION**

A simple model that explains the formation and the topological features of an EP degeneracy is given by a two-mode model. The effective Hamiltonian that describes such system is given by:

$$H_2 = \begin{pmatrix} V - i\gamma & T \\ T & V + i\gamma \end{pmatrix} \tag{1}$$

in which $V$ is the resonant frequency of each mode, $\gamma$ is the corresponding gain/loss coefficient describing attenuation and amplification mechanisms associated with each mode, and $T$ is the coupling between the two modes. It is easy to show that such Hamiltonian commutes with the joint parity-time-symmetric operator, and therefore we refer to it as a PT-symmetric Hamiltonian. The corresponding eigenvalues $\lambda_\pm = V \pm \sqrt{T^2 - \gamma^2}$ are shown in Fig. 1 and they are characterized by a transition from real valued ($|T| > |\gamma|$) to complex conjugate pair ($|T| < |\gamma|$). The transition point ($|T| = |\gamma| \neq 0$) has the square-root characteristic feature of an EP degeneracy (see Figs. 1(d)-(e)). The former domain is known as the exact PT-symmetric phase and the eigenvectors are also the eigenvectors of the PT-operator. The other domain is known as the broken PT-symmetric phase. In this case, the eigenvectors are not any more the eigenvectors of the PT-operator. In the upper plots of Fig. 1 (Figs. 1(a)-(c), in which the Riemann surfaces are identical), we report a parametric evolution of the real part of the eigenvalues vs. $T$ and $\gamma$. Fig. 1(d) shows the real and imaginary parts of the eigenvalues when $\gamma = 5$, i.e., the intersection between the Riemann surfaces and the red plane in Fig. 1(a). In this plot, we can clearly see the formation of two EPs at $T = \pm|\gamma| = \pm 5$ demonstrating the square-root characteristic of the EP. Similarly, when $T$ is fixed to a certain value, for example $T = 5$ as shown in Fig. 1(e), two EPs can be found at $\gamma = \pm|T| = \pm 5$. The cross-section of the Riemann surfaces with the specific plane corresponding to $\gamma = 0$ where the Hamiltonian $H_2$ is Hermitian is also shown in Fig. 1(f). In this specific case, the system develops a Dirac point degeneracy characterized by a linear dispersion.

In this study, we consider a more complicated scenario than the one discussed above where a PT-symmetric dimer is imbedded in a periodic metastructure as a pair of defects and analyze the emergence of the EPs in the spectrum of such system by performing steady-state dynamic analyses in commercial finite element software ABAQUS Simulia. The metastructure has $N$ cross beams—each beam has an identical total length of $28 \ \mu m$ and an interval of $10 \ \mu m$—on each (P-symmetric) side coupled via a long horizontal beam with $20 \ \mu m$ ledges at both ends. Fig. 2(a) illustrates an example of an N6 (representing $N = 6$) periodic metastructure with 6 cross beams on each side of the mirror plane (represented by transparent yellow), which can be paired and labeled as ±1, ±2, … ±6 starting from the middle, i.e., the mirror plane. Considering its application in MEMS-scale devices, we model the material to be silicon nitride (Young's modulus: 290 GPa, density: 3000 kg/m3, Poisson's ratio: 0.27) and utilize 3-node quadratic Timoshenko beam elements with a uniform rectangular cross-section ($2\ \mu m$ widths, $0.2\ \mu m$ thickness) throughout the model. Each cross beam is discretized with forty elements while the coupling beam segment at each interval has twenty elements. After fixing both ends, we harmonically excited the metastructure at the mid-point of its left ledge by a transverse force and measured the sinusoidal displacement response at the corresponding right side. COMSOL Multiphysics has only been used to calculate the band structure of the system because of its simplicity in implementation compared with ABAQUS Simulia and we chose the same Timoshenko beam elements and verified the consistency between the displacement responses in both finite element software. Equal magnitudes of the structural anti-damping and damping rates represent the gain and loss mechanisms in the system during simulations, respectively.

To identify the spectral location of EP emergence in a periodic metastructure with a PT-symmetric defect, we investigate the displacement spectrum from $0\sim30$ MHz in an N21 (representing $N = 21$) Hermitian periodic structure in Fig. 2($b$), where five bands are indicated in five different colors. When we introduce gain (loss) to the left (right) half of the system and increase their intensity, almost all the modes in each band shown in Fig. 2($b$) start to degenerate and form EPs. Fig. 3($a$) illustrates such trend in band 4, and the disappearing peaks are evidence of the emergence of EPs as well as the damping of the modes. For a system with only one PT-symmetric defect, however, the modes coalesce and form EPs only at the band edge regions (indicated by grey arrows on the x-axis of Fig. 2($b$)) due to the higher density of modes in band edges. Such behavior can be seen in Fig. 4(*a*), where the limited modes near band edges become degenerate when $\gamma$ is increased. Fig. 4(*b*) shows the formation of a typical EP in the metastructure with a single PT-symmetric defect, and Fig. 4(*c*) exhibits the variation of its corresponding real and imaginary parts of the frequency. Fig. 4(*d*) shows a square-root behavior typical of order-two EP, where $\Delta_{EP}$ is the frequency difference between the corresponding two modes which degenerate to form the EP at the critical gain/loss intensity $\gamma_{EP}$. The aforementioned behavior is similar to that of an EP formed in an entire PT-symmetric structure (see Figs. 3(*b*)-(*d*)).

We further investigate the relationship between the emerging number of EPs and the location of the PT-symmetric defect. Figs. 5(*a*)-(*b*) show that the number of EPs in each of the five bands decreases as the distance between the two resonators that make up the PT-symmetric defect, i.e., the defect position index, is increased from ±1 to ±21. This decreasing EP number density is clearly observable in high frequency bands than low frequency ones because of the higher EP density in high frequency bands. Note that an appropriate comparison among different bands requires that the maximum introduced gain/loss intensity $\gamma$ of the entire band is normalized to its average level spacing as it affects the critical gain/loss intensity for EP formation depending on the initial split between the modes [25]. Here, we show the variation of EP numbers whose critical non-Hermitian strength $\gamma_{EP} \leq 0.01 \cdot \Delta_{ave}$ (see Fig. 5(*a*)) and $\gamma_{EP} \leq 0.001 \cdot \Delta_{ave}$ (see Fig. 5(*b*)), respectively. This observation implies that we can utilize the defect position as another effective control parameter to tailor the emergence of EPs: the stronger the coupling between the PT-symmetric defects is, the higher the number density of the EPs within a band under the same given value of the intensity $\gamma$ is. We develop a model based on coupled-mode-theory (CMT) to examine this relation between defect position and the number density of the EPs in a sufficiently larger system. Its Hamiltonian can be described by:

$$H = \left(\sum_{n=-N}^{-1}|n\rangle\widetilde{V_n}\langle n| + \sum_{n=-N}^{-2}|n+1\rangle t\langle n| + c.c.\right) + \left(\sum_{n=1}^{N}|n\rangle\widetilde{V_n}\langle n| + \sum_{n=1}^{N-1}|n+1\rangle t\langle n| + c.c.\right) + (|1\rangle t\langle -1| + c.c.)$$

(2)

where $|n\rangle$ is the basis of the local mode, $t = -1$ is the coupling constant between neighboring potentials, $\widetilde{V_n} = i\gamma\left(-\delta_{n,-d} + \delta_{n,d}\right)$ stands for the coupled on-site potentials with a pair of defects at sites $n = \pm d$ that respect PT-symmetry ($d = 1,2,3 \dots N$ is defect position). Fig. 5(c) shows a clear inverse relation between the number density of EPs and the defect position calculated from the CMT model. The number of EPs ($N_{EP}$), the defect position $d$, and the half size $N$ of the system are related by:

$$N_{EP} = N + 1 - d$$

(3)

Note that the CMT model does not differentiate the PT-symmetric response between the lower and higher frequency bands because of its generic nature.

We also notice that the defect position influences the critical non-Hermitian strength $\gamma_{EP}$ of the first EP in each band. The first EP in each band originates from two modes whose level spacing in the Hermitian spectra is the smallest among all other mode pairs within that band. Fig. 6(a) shows the results from the FE models, where the $\gamma_{EP}$ first decreases and then increases as the defect position index increases encompassing a minimum. This finding unveils another utility of specifically positioning the PT-symmetric defect in a metastructure: it affects the critical gain/loss intensity required to induce a phase transition in the metastructure going from the exact phase to the broken phase. This peculiar phenomenon can also be predicted by CMT models, for example see the red curve in Fig. 6(b) corresponding to N50 ($N = 50$) system. To further elucidate this behavior, we use a TISOP theory. The Hamiltonian of the system $H$ can be treated as:

$$H = H_0 + i\gamma V \tag{4}$$

where $H_0$ is the unperturbed matrix with eigenvalues $E^{(0)}$ and the corresponding eigenvector $\phi = |\phi^{(n)}\rangle$ [39], $V$ is the perturbation with diagonal elements $V_n = -\delta_{n,-d} + \delta_{n,d}$ where $|n| = 1,2,\ldots N$:

$$H_0 = (\sum_{n=-N}^{-2}|n+1\rangle t\langle n| + \sum_{n=1}^{N-1}|n+1\rangle t\langle n| + |1\rangle t\langle -1|) + c.c. \tag{5}$$

$$V = \sum_n |n\rangle V_n \langle n| \tag{6}$$

$$E_k^{(0)} = 2t \cdot \cos\frac{k\pi}{2N+1} \xrightarrow{R=\pi/(2N+1)} 2t \cdot \cos(kR) \tag{7}$$

$$\phi_k^{(n)} = \frac{t}{|t|} \cdot \sqrt{\frac{2}{2N+1}} \cdot \sin\left(\frac{kn\pi}{2N+1}\right) \xrightarrow{R=\pi/(2N+1)} \frac{t}{|t|} \cdot \sqrt{\frac{2}{2N+1}} \cdot \sin(knR) \tag{8}$$

The $k^{th}$ eigenvalue $E_k$ of $H$ can be written as a power series:

$$E_k = E_k^{(0)} + i\gamma E_k^{(1)} + (i\gamma)^2 E_k^{(2)} + \cdots \tag{9}$$

where $E_k^{(1)}$ and $E_k^{(2)}$ are the first and second order corrections to $E_m$ (see supplementary material for details):

$$E_k^{(1)} = \langle\phi_k|V|\phi_k\rangle \equiv \langle k|V|k\rangle = -\frac{2}{2N+1}\left[sin^2(kaR) - sin^2(kbR)\right] \tag{10}$$

$$E_k^{(2)} = \sum_{n\neq k}\frac{|\langle k|V|n\rangle|^2}{E_k^{(0)}-E_n^{(0)}} = \frac{2}{t(2N+1)^2}\sum_{n\neq k}\frac{[sin(kaR)\cdot sin(naR) - sin(kbR)\cdot sin(nbR)]^2}{cos(kR)-cos(nR)} \tag{11}$$

where $a = N + 1 - d$ and $b = N + d$ are the locations of defects. The real part of $E_m$ can be represented as:

$$Real(E_k) \approx E_k^{(0)} + (i\gamma)^2 E_k^{(2)} = E_k^{(0)} - \gamma^2 E_k^{(2)} \tag{12}$$

By setting the real parts of the two modes we studied ($E_1$ and $E_2$) equal, we obtain the expression for the corresponding critical gain/loss intensity $\gamma_{EP}$:

$$\gamma_{EP} = \left[ \frac{t^2(2N+1)^2 \cdot [cos(R) - cos(2R)]}{\sum_{n \neq 1} \frac{[sin(aR) \cdot sin(naR) - sin(bR) \cdot sin(nbR)]^2}{cos(R) - cos(nR)} - \sum_{n \neq 2} \frac{[sin(2aR) \cdot sin(naR) - sin(2bR) \cdot sin(nbR)]^2}{cos(2R) - cos(nR)}} \right]^{1/2} \tag{13}$$

Its variation is shown in Fig. 6(b) (blue curve), revealing an agreement with the CMT model: both of them have the same increasing and decreasing trends, and the critical point for $\gamma_{EP}$ (i.e., the location where $\gamma_{EP}$ reaches its minimum) is around 17 in an N50 model. When the defect position locates at the site of this critical point, the threshold value of gain/loss intensity $\gamma_{EP}$ will be the smallest for making those two modes degenerate and form an EP, implying the smallest needed energy to realize the PT-phase transition in the corresponding band.

To further explore the relationship between the critical point of $\gamma_{EP}$ and the size of the system, we compare the calculation results from both the CMT model and TISOP theory. Fig. 6(c) indicates a linear relationship between the critical point of $\gamma_{EP}$ and the half size $N$ of the system. The fitting slopes for the CMT model (red line) and the TISOP theory calculation (blue line) are 0.312 and 0.333, respectively, exhibiting great agreement again.

It is natural to assume that the EP degeneracy will be easier enforced for nearby levels. Therefore, we expect that $\Delta_0$ and $\gamma_{EP}$, i.e., the initial frequency split between the modes which will degenerate to form EP and the critical gain/loss intensity required to form that EP, will be linearly related to one another. Indeed, such fashion has been confirmed in the case of elastodynamics structures with fractal spectra [25], and we have also confirmed that this linear relationship persists. We compared the relationship between $\Delta_0$ and $\gamma_{EP}$ among the EPs in the third band of N21 periodic metastructure in Fig. 7(a) by using different colors to represent different systems: the entire PT-symmetric system (black), systems with PT-symmetric defects at position ±3 (red), ±5 (blue), and ±7 (green). The fitting slopes verify the linear relationship. In the system with PT-symmetric defects, however, the fitting slopes contain a slightly larger offset when two defects are too close, which can be observed in Fig. 7(b), showing the variation of the fitting slope of EPs with respect to the defect position. For the first five bands in the N21 FE model of periodic metastructure, the slopes come close to 1 when the defect position increases to 5. For the N50 CMT model, the defect position index should be at least around 10 for obtaining a linear slope. It should be noted that the FE model simulations require significant computational cost, and the small number of EP points investigated may affect the accuracy of the slopes especially when the defect position index is large.

**CONCLUSIONS**

In summary, we designed a periodic metastructure and investigated in a steady-state dynamic FE model and CMT-based mathematical model approach the emergence of EPs when a pair of defects that enforce a PT-symmetry to the whole structure are introduced at different positions. When the interval between the pair of PT-symmetric defects increases, the number of EPs that emerge in all five bands below 30 MHz decreases. Furthermore, we revealed the variation of the threshold value of the gain/loss intensity $\gamma_{EP}$ of the first EP in each band by utilizing both the FE model and CMT-based mathematical model and demonstrated the relationship between its critical point (i.e., the defect position where $\gamma_{EP}$ reaches its minima) and system size by CMT model and TISOP theory. Our findings show that the PT-symmetric defect position can be exploited to effectively tailor and control the emergence and the density of EPs in an elastodynamic metastructure paving the way for their convenient experimental implementations in the elastodynamic metastructures and metamaterials.

**ACKNOWLEDGMENT**

This work was supported by the Dynamics, Controls, and System Dynamics (DCSD) Program of the National Science Foundation (NSF) under the awards CMMI-1925530 and CMMI-1925543.

**REFERENCES**

[1]    Bender, C. M., and Boettcher, S., 1998, "Real Spectra in Non-Hermitian Hamiltonians Having PT Symmetry," Physical Review Letters, **80**(24), pp. 5243–5246.

[2]    Bender, C. M., Boettcher, S., and Meisinger, P. N., 1999, "PT-Symmetric Quantum Mechanics," Journal of Mathematical Physics, **40**(5), pp. 2201–2229.

[3]    Bender, C. M., Brody, D. C., and Jones, H. F., 2002, "Complex Extension of Quantum Mechanics," Physical Review Letters, **89**(27), pp. 1–4.

[4]    Bender, C. M., 2007, "Making Sense of Non-Hermitian Hamiltonians," Reports on Progress in Physics, **70**(6), pp. 947–1018.

[5]    Bendix, O., Fleischmann, R., Kottos, T., and Shapiro, B., 2009, "Exponentially Fragile PT Symmetry in Lattices with Localized Eigenmodes," Physical Review Letters, **103**(3), p. 030402.

[6]    El-Ganainy, R., Makris, K. G., Christodoulides, D. N., and Musslimani, Z. H., 2007, "Theory of Coupled Optical PT-Symmetric Structures," Optics Letters, **32**(17), p. 2632.

[7]    Makris, K. G., El-Ganainy, R., Christodoulides, D. N., and Musslimani, Z. H., 2008, "Beam Dynamics in PT Symmetric Optical Lattices," Physical Review Letters, **100**(10), p. 103904.

[8]    Guo, A., Salamo, G. J., Duchesne, D., Morandotti, R., Volatier-Ravat, M., Aimez, V., Siviloglou, G. A., and Christodoulides, D. N., 2009, "Observation of PT-Symmetry Breaking in Complex Optical Potentials," Physical Review Letters, **103**(9), pp. 1–4.

[9]    Rüter, C. E., Makris, K. G., El-Ganainy, R., Christodoulides, D. N., Segev, M., and Kip, D., 2010, "Observation of Parity-Time Symmetry in Optics," Nature Physics, **6**(3), pp. 192–195.

[10]   El-Ganainy, R., Makris, K. G., Khajavikhan, M., Musslimani, Z. H., Rotter, S., and Christodoulides, D. N., 2018, "Non-Hermitian Physics and PT Symmetry," Nature Physics, **14**(1), pp. 11–19.

[11]   Zhu, B., Lü, R., and Chen, S., 2014, "PT Symmetry in the Non-Hermitian Su-Schrieffer-Heeger Model with Complex Boundary Potentials," Physical Review A - Atomic, Molecular, and Optical Physics, **89**(6), pp. 1–6.

[12]   Lin, Z., Ramezani, H., Eichelkraut, T., Kottos, T., Cao, H., and Christodoulides, D. N., 2011, "Unidirectional Invisibility Induced by PT-Symmetric Periodic Structures," Physical Review Letters, **106**(21), p. 213901.

[13]   Bittner, S., Dietz, B., Günther, U., Harney, H. L., Miski-Oglu, M., Richter, A., and Schäfer, F., 2012, "PT Symmetry and Spontaneous Symmetry Breaking in a Microwave Billiard," Physical Review Letters, **108**(2), p. 024101.

[14]   Doppler, J., Mailybaev, A. A., Böhm, J., Kuhl, U., Girschik, A., Libisch, F., Milburn, T. J., Rabl, P., Moiseyev, N., and Rotter, S., 2016, "Dynamically Encircling an Exceptional Point for Asymmetric Mode Switching," Nature, **537**(7618), pp. 76–79.

[15]   Schindler, J., Lin, Z., Lee, J. M., Ramezani, H., Ellis, F. M., and Kottos, T., 2012, "PT-Symmetric Electronics," Journal of Physics A: Mathematical and Theoretical, **45**(44), p. 444029.

[16]   Assawaworrarit, S., Yu, X., and Fan, S., 2017, "Robust Wireless Power Transfer Using a Nonlinear Parity-Time-Symmetric Circuit," Nature, **546**(7658), pp. 387–390.

[17]   Zhu, X., Ramezani, H., Shi, C., Zhu, J., and Zhang, X., 2014, "PT -Symmetric Acoustics," Physical Review X, **4**(3), p. 031042.

[18]   Fleury, R., Sounas, D., and Alù, A., 2015, "An Invisible Acoustic Sensor Based on Parity-Time Symmetry," Nature Communications, **6**(1), p. 5905.

[19]   Aurégan, Y., and Pagneux, V., 2017, "P T -Symmetric Scattering in Flow Duct Acoustics," Physical Review Letters, **118**(17), p. 174301.

[20]   Shi, C., Dubois, M., Chen, Y., Cheng, L., Ramezani, H., Wang, Y., and Zhang, X., 2016, "Accessing the Exceptional Points of Parity-Time Symmetric Acoustics," Nature Communications, **7**(1), p. 11110.

[21]   Achilleos, V., Theocharis, G., Richoux, O., and Pagneux, V., 2017, "Non-Hermitian Acoustic Metamaterials: Role of Exceptional Points in Sound Absorption," Physical Review B, **95**(14), p. 144303.

[22]   Ding, K., Ma, G., Xiao, M., Zhang, Z. Q., and Chan, C. T., 2016, "Emergence, Coalescence, and Topological Properties of Multiple Exceptional Points and Their Experimental Realization," Physical Review X, **6**(2), p. 021007.

[23]   Thevamaran, R., Branscomb, R. M., Makri, E., Anzel, P., Christodoulides, D., Kottos, T., and Thomas, E. L., 2019, "Asymmetric Acoustic Energy Transport in Non-Hermitian Metamaterials," J Acoust Soc Am, **146**(1), pp. 863–872.

[24]   Domínguez-Rocha, V., Thevamaran, R., Ellis, F. M., and Kottos, T., 2020, "Environmentally Induced Exceptional Points in Elastodynamics," Physical Review Applied, **13**(1), p. 014060.

[25]   Fang, Y., Kottos, T., and Thevamaran, R., 2021, "Universal Route for the Emergence of Exceptional Points in PT-Symmetric Metamaterials with Unfolding Spectral Symmetries," New Journal of Physics, **23**(6), p. 63079.

[26]   Kononchuk, R., Cai, J., Ellis, F., Thevamaran, R., and Kottos, T., 2022, "Enhanced Signal-to-Noise Performance of EP-Based Electromechanical Accelerometers."

[27]   Gupta, A., Kurnosov, A., Kottos, T., and Thevamaran, R., 2022, "Boosting of Purcell Enhancement Factor near Exceptional Point Degeneracies in an Elastodynamic Metamaterial."

[28]   Lustig, B., Elbaz, G., Muhafra, A., and Shmuel, G., 2019, "Anomalous Energy Transport in Laminates with Exceptional Points," Journal of the Mechanics and Physics of Solids, **133**, p. 103719.

[29]   Hou, Z., and Assouar, B., 2018, "Tunable Elastic Parity-Time Symmetric Structure Based on the Shunted Piezoelectric Materials," Journal of Applied Physics, **123**(8).

[30]   Rosa, M. I. N., Mazzotti, M., and Ruzzene, M., 2021, "Exceptional Points and Enhanced Sensitivity in PT-Symmetric Continuous Elastic Media," Journal of the Mechanics and Physics of Solids, **149**.

[31]    Bludov, Y. V., Hang, C., Huang, G., and Konotop, V. V., 2014, "PT-Symmetric Coupler with a Coupling Defect: Soliton Interaction with Exceptional Point," Optics Letters, **39**(12), p. 3382.

[32]    Abdullaev, F. K., Brazhnyi, V. A., and Salerno, M., 2013, "Scattering of Gap Solitons by PT-Symmetric Defects," Physical Review A - Atomic, Molecular, and Optical Physics, **88**(4), pp. 1–9.

[33]    Zhang, X., Chai, J., Huang, J., Chen, Z., Li, Y., and Malomed, B. A., 2014, "Discrete Solitons and Scattering of Lattice Waves in Guiding Arrays with a Nonlinear PT - Symmetric Defect," Optics Express, **22**(11), p. 13927.

[34]    Musslimani, Z. H., Makris, K. G., El-Ganainy, R., and Christodoulides, D. N., 2008, "Optical Solitons in PT Periodic Potentials," Physical Review Letters, **100**(3), p. 030402.

[35]    Regensburger, A., Miri, M. A., Bersch, C., Näger, J., Onishchukov, G., Christodoulides, D. N., and Peschel, U., 2013, "Observation of Defect States in PT-Symmetric Optical Lattices," Physical Review Letters, **110**(22), p. 223902.

[36]    Jin, L., Wang, P., and Song, Z., 2017, "Su-Schrieffer-Heeger Chain with One Pair of PT-Symmetric Defects," Scientific Reports, **7**(1), pp. 1–9.

[37]    Mostafavi, F., Yuce, C., Maganã-Loaiza, O. S., Schomerus, H., and Ramezani, H., 2020, " Robust Localized Zero-Energy Modes from Locally Embedded  PT - Symmetric Defects ," Physical Review Research, **2**(3), pp. 1–7.

[38]    Dmitriev, S. v., Suchkov, S. v., Sukhorukov, A. A., and Kivshar, Y. S., 2011, "Scattering of Linear and Nonlinear Waves in a Waveguide Array with a PT-Symmetric Defect," Physical Review A - Atomic, Molecular, and Optical Physics, **84**(1), pp. 1–5.

[39]    Izrailev, F. M., Kottos, T., and Tsironis, G. P., 1996, "Scaling Properties of the Localization Length in One-Dimensional Paired Correlated Binary Alloys of Finite Size," Journal of Physics: Condensed Matter, **8**(16), pp. 2823–2834.

defects. The formation of a typical EP found in the dashed cyan boxed region in band 4 is shown in Figs. 3-4.

Fig. 3    (a) The variation of the displacement spectra in band 4 with respect to the gain/loss intensity parameter $\gamma$ in an entirely PT-symmetric N21 periodic metastructure. (b) The formation of a typical EP found in the location marked by a cyan box in (a) as well as Fig. 2(b). (c) Corresponding real (red lines) and imaginary parts (blue lines) of mode frequency, where solid and hollow dots denote two different eigenvalues, and (d) a log-log plot of the frequency difference $\Delta_{EP}$ vs $1 - \gamma/\gamma_{EP}$ near this EP. The red fit line has a slope of 0.46, reflecting the square-root characteristic typical of an order two EP.

Fig. 4    (a) The variation of the displacement spectra in band 4 with respect to the gain/loss intensity parameter $\gamma$ from an N21 periodic metastructure with a pair of PT-symmetric defects on cross beams ±3. (b) The formation of a typical EP found in the location marked by a cyan box in (a) as well as Fig. 2(b). (c) Corresponding real (red lines) and imaginary parts (blue lines) of mode frequency, where solid and hollow dots denote two different eigenvalues, and (d) a log-log plot of the frequency difference $\Delta_{EP}$ vs $1 - \gamma/\gamma_{EP}$ near this EP. The red fit line has a slope of 0.51, reflecting the square-root characteristic typical of an order two EP.

Fig. 5        Variation of the number of EPs in each band with respect to the defect

position in N21 periodic metastructure when the critical non-Hermitian strength  (a)  $\gamma_{EP} \leq 0.01 \cdot \Delta_{ave}$  and  (b)  $\gamma_{EP} \leq 0.001 \cdot \Delta_{ave}$ ,  and  (c) coupled-mode-theory (CMT) models for different sizes ( $N$ ) of the metastructure.

Fig. 6        Variation of $\gamma_{EP}$ with respect to the defect position for two modes with the smallest level spacing in one band in (a) N21 FE model of periodic metastructure and (b) N50 CMT model as well as TISOP theory calculation.

(c) The relationship between the critical point of $\gamma_{EP}$ and the half size $N$ of

the system. The red and blue lines are linear fittings for the CMT model

(slope 0.312) and TISOP theory calculation (slope 0.333), respectively.

Fig. 7        (a) The relationship between $\Delta_0$ and $\gamma_{EP}$ from EPs in the third band of N21

PT-symmetric periodic metastructures with no defects (black), defect

position at 3 (red), 5 (blue), and 7 (green). Lines are the linear fittings for

the corresponding systems, and the slopes are shown in the legend. (b)

Variation of the fitting slopes of $\Delta_0$ and $\gamma_{EP}$ with respect to the defect

position in the N50 CMT model and the first five bands from the N21

periodic metastructure. For FE analysis, $\gamma_{EP} \leq 0.001 \cdot \Delta_{ave}$ for those EPs.

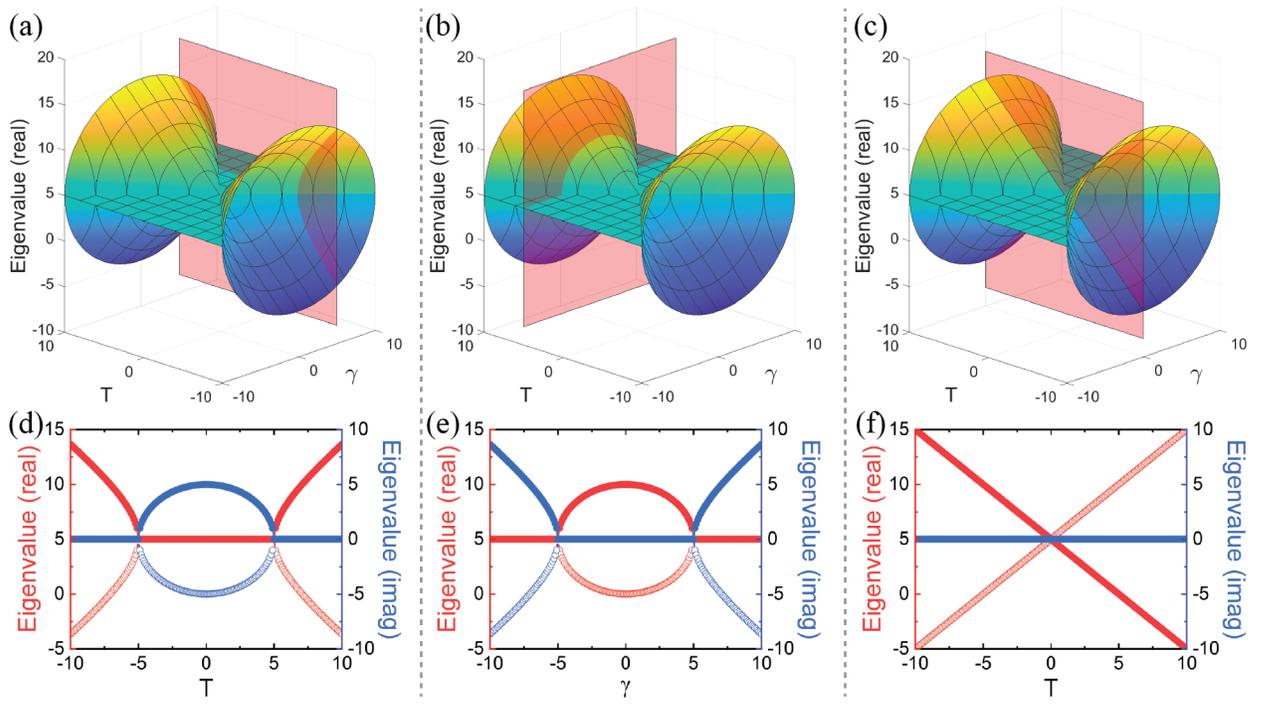

(a)

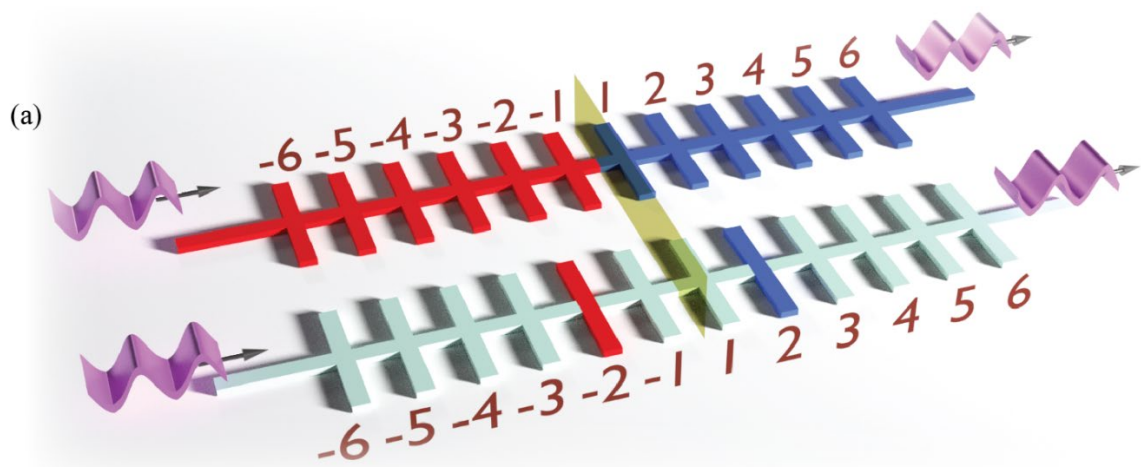

(b)

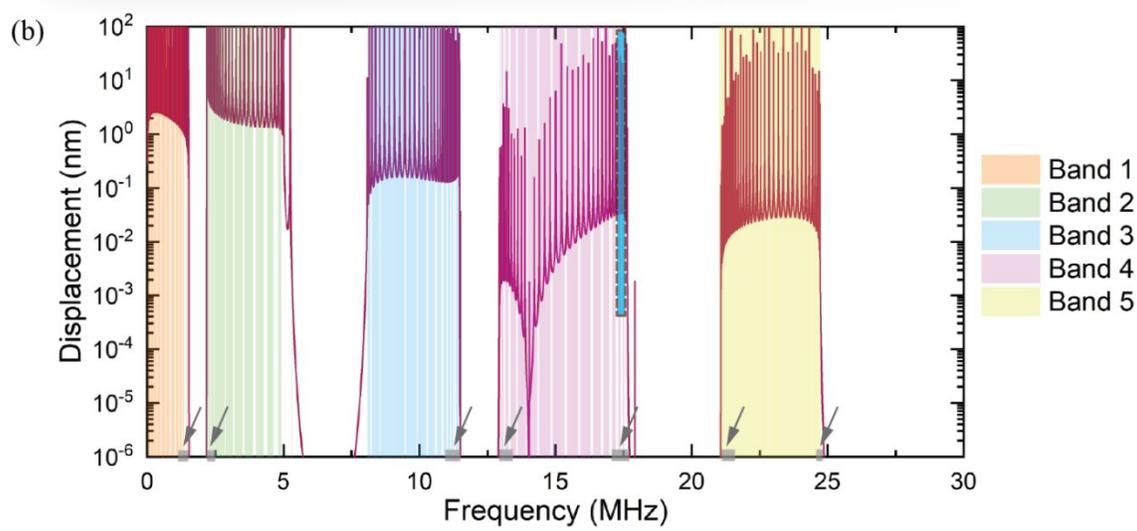

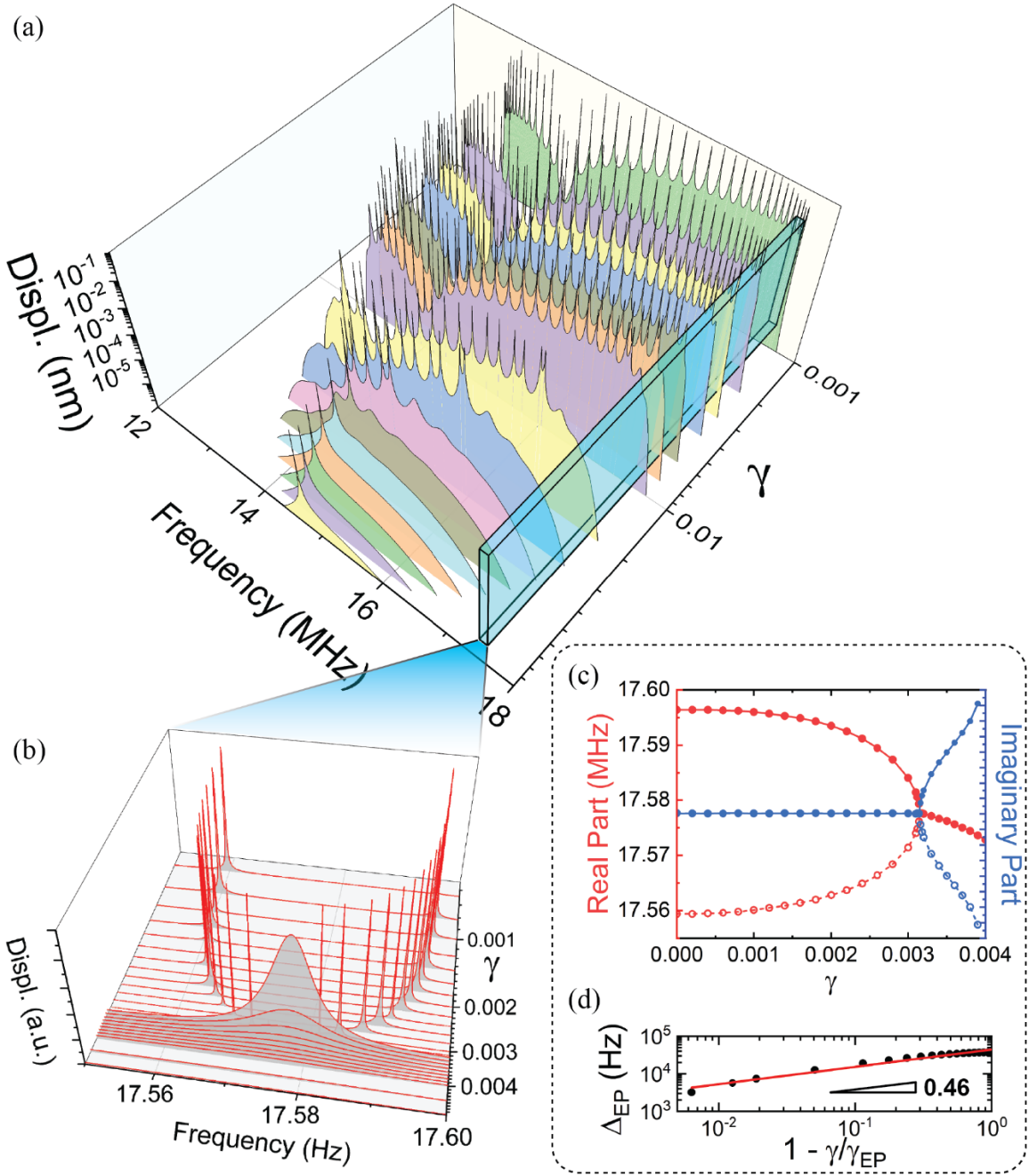

(a)

(b)

(c)

(d)

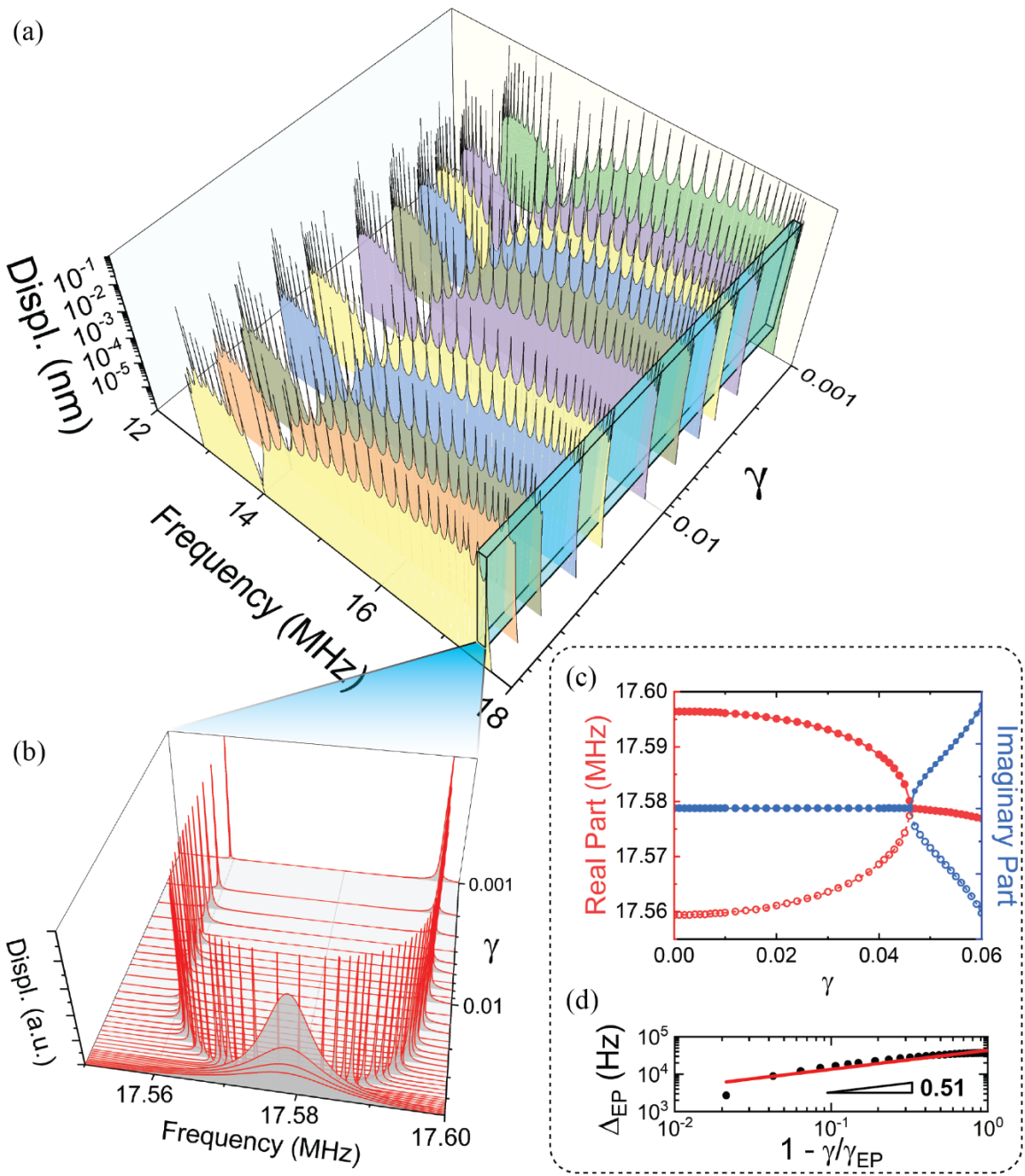

(a)

(b)

(c)

(d)

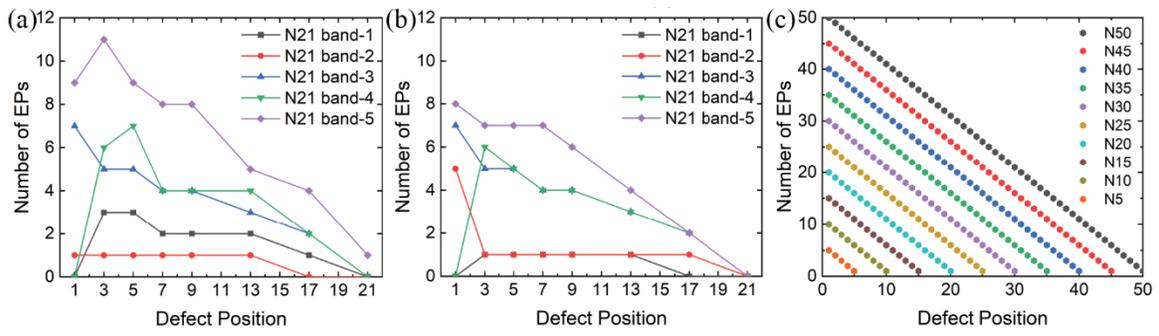

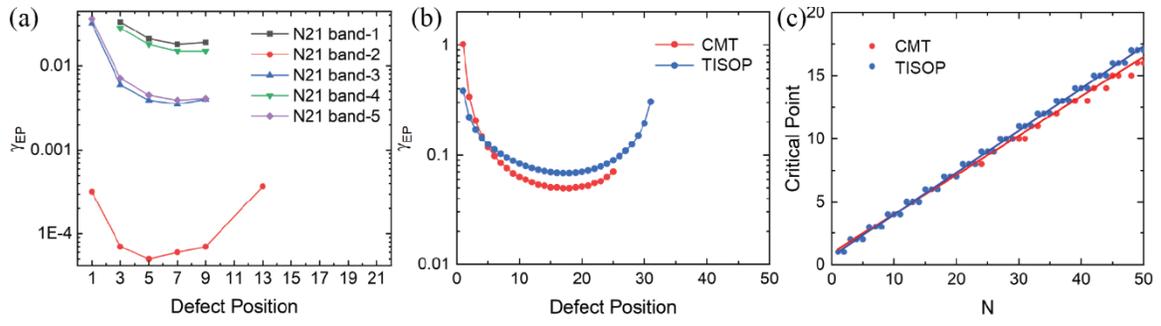

(a) 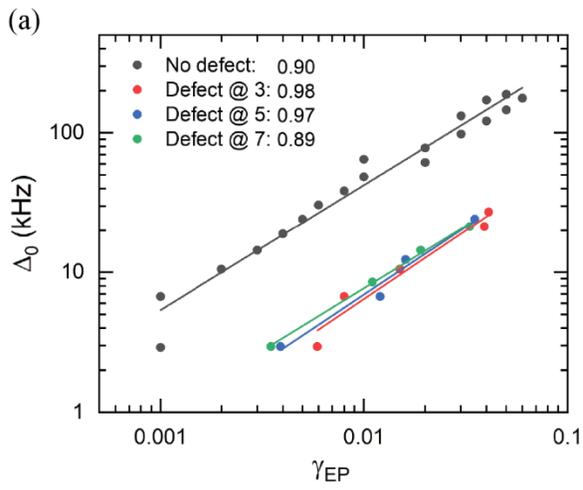

(b) 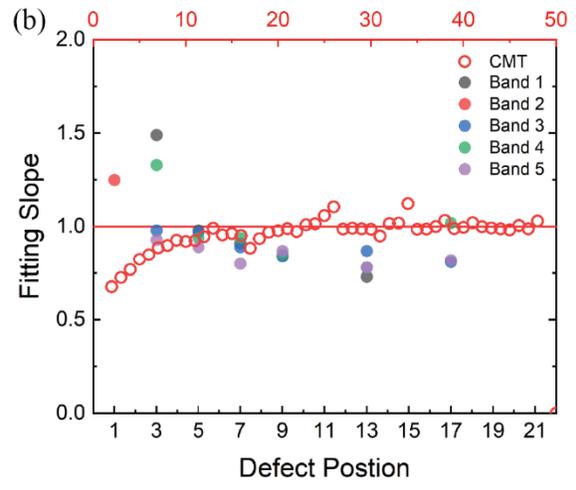